\documentclass{ws-procs9x6}

\begin{document}

\title{On the modulational stability of Gross-Pittaevskii type equations in 1+1 dimensions}

\author{Z. Rapti}

\address{Department of Mathematics and Statistics, 
University of Massachusetts, Amherst MA 01003-4515, USA\\
E-mail: rapti@math.umass.edu}

\author{P.~G. Kevrekidis}

\address{Department of Mathematics and Statistics, 
University of Massachusetts, Amherst MA 01003-4515, USA\\
E-mail: kevrekid@math.umass.edu
}

\author{V.~V. KONOTOP }

\address{Centro de F\'{\i}sica 
de Mat\'eria Condensada, Universidade de Lisboa, Av. Prof. Gama Pinto, 2, Lisboa 1649-003, Portugal\\ 
E-mail: konotop@cii.fc.ul.pt}

\maketitle

\abstracts{The modulational stability of the
nonlinear Schr{\"o}dinger (NLS) equation 
is examined in the cases with  linear and quadratic
external
potential. This study is motivated by recent experimental
studies in the context of matter waves in Bose-Einstein condensates (BECs).
The linear case can be examined by means of the Tappert transformation
and can be mapped to the NLS in the appropriate (constant acceleration)
frame. The quadratic case can be examined by using a lens-type 
transformation that converts it into a regular NLS with an additional
linear growth term.}

\section{Introduction}

Intensive studies of Bose-Einstein condensates (BECs) \cite{review} have 
drawn much attention to nonlinear excitations in them. Recent experiments 
have revealed the existence of bright \cite{bright} and 
dark \cite {dark,dexp2,nsbec} solitons in BECs, as well as topological 
structures, such as vortices \cite{vortex} and vortex lattices \cite{vl}.
An interesting question concerns how such solitary wave structures may
arise in this novel context of matter waves in BECs. It is well-known
that the dynamics of the wavefunction in BEC is described (at the
mean field level, which is an increasingly accurate description, 
the closer the system is to the
zero temperature limit) by the 
Gross-Pittaevskii (GP) equation which is a variant of well-known
nonlinear Schr{\"o}dinger (NLS) equation \cite{sulem}. In the NLS 
context, perhaps the most standard mechanism through which solitons
and solitary wave structures appear is through the activation of the
modulational instability (MI) of plane waves. 

The MI is a general feature of discrete as well as continuum nonlinear wave 
equations.
Its demonstrations span a diverse set of disciplines ranging from 
fluid dynamics \cite{benjamin67} (where it is usually referred to as the 
Benjamin-Feir instability) 
and nonlinear optics \cite{ostrovskii69} to plasma physics \cite{taniuti68}.
One of the early contexts in which its significance was appreciated was 
the linear stability analysis of deep water waves. 

The MI has been examined recently in the context of optical 
lattices in BECs both in one-dimensional and quasi-one dimensional
systems, as well as in multiple dimensions. In the former
case, it has been predicted theoretically \cite{konotop,smerzi}
and verified experimentally \cite{cattal,kasevich} to 
lead to destabilization of plane waves, and in turn to
delocalization in momentum space (equivalent to localization in position
space, and the formation of solitary wave structures).

In the present contribution, we discuss the MI conditions 
for the continuous NLS equation (or equivalently
for the GP equation) in (1+1) dimensions (1 spatial and 1 temporal)
\begin{eqnarray}
i u_t + u_{xx} + s |u|^2 u + V(x) u=0.
\label{req1}
\end{eqnarray}
$u$ in this equation describes the slow envelope complex field
dynamics (modulating the fast oscillatory dynamics). The subscripts
denote partial derivatives with respect to the index, $s 
\in \{1,-1\}$ illustrates the focusing ($+1$) or defocusing ($-1$)
nature of the nonlinearity, while $V(x)$ is the external potential.

In section 2.1, we review briefly the results in the absence of the
potential (e.g., for $V(x) \equiv 0$). In section 2.2, we examine
the case of a linear potential:
\begin{eqnarray}
V(x)=-\alpha x,
\label{req2}
\end{eqnarray}
which is relevant in experimental situations with gravitational
\cite{kasevich1} (and potentially also electrostatic) fields.
%This context
%where the nonlinearities are of the 
%following two kinds:
%\begin{equation}
%iu_{t}+u_{xx}+|u|^2 u-2 \alpha x u=0  
%\end{equation} 
%and 
In section 2.3, we examine the quadratic potential of the general form:
\begin{eqnarray}
V(x)=-k(t) x^2,
\label{req3}
\end{eqnarray}
which is relevant to contexts in which the (magnetic) trap is 
strongly confined
in the 2 directions, while it is much shallower in the third one \cite{review}.
The prefactor $k(t)$ is typically fixed in current experiments, but adiabatic
changes in the strength (and in fact even the location of the center) 
of the trap are experimentally feasible, hence we examine the more general
time-dependent case.
Finally, in section 3 we summarize our findings and conclude.

%\begin{equation}
%iu_{t}+u_{xx}+|u|^2 u-k(t) x^2 u=0   
%\end{equation}
%We analyze the stability condition of these equations.
\section{Analytical Results}
\subsection{No potential}
We start by recalling the results for the modulational stability of
the NLS (\ref{req1}) without an external potential, i.e. for $V(x)\equiv 0$:
%\begin{equation}
%iu_{t}+u_{xx}+ s |u|^2 u=0               
%\label{req4}
%\end{equation}
To this end we look for perturbed plane wave solutions of the form
\begin{equation}
u(x,t)=(\phi+\epsilon b) \exp[i ((q x-\omega t) +\epsilon \psi(x,t))]
\label{req5}
\end{equation}
and analyze the $O(\epsilon)$ terms as
\begin{eqnarray}
b(x,t)=b_{0} \exp(i\beta(x,t)), \quad
\psi(x,t)=\psi_{0} \exp(i\beta(x,t)).
\label{req6}
\end{eqnarray} 
Using
%\begin{equation}
$
\beta(x,t)=Q x-\Omega t,
$
%\label{req7}
%\end{equation}
the dispersion relation connecting the  wavenumber $Q$ and
frequency $\Omega$ of the perturbation (see e.g. \cite{sulem})
\begin{equation}
(-\Omega+2 q Q)^{2}=Q^{2} (Q^{2}-2 s \phi^{2})
\label{req8}
\end{equation}
is obtained.

This implies that the instability region for the NLS in the absence
of an external potential, appears for perturbation wavenumbers 
$Q^2<{2 s} \phi^2$, 
and in particular {\it only for focusing nonlinearities}
(to which we will restrict our study from this point onwards).

\subsection{Linear Potential}

The case of a linear potential is relevant to any context of a costant
external field (gravitational \cite{kasevich1} and electric ones being
among the prominent such examples). In this setting, the NLS is well-known
to maintain its integrable character \cite{liu}. Hence, in some sense, we
expect that the modulational results/conditions will not be modified
in this case.

The simplest way to illustrate that is by means of the 
``Tappert transformation'' \cite{liu} $(\xi=x+\alpha t^2)$
\begin{eqnarray}
u(x,t)=v(\xi,t) \exp(-i \alpha x t-\frac{1}{3} i \alpha^2 t^3)
\label{req9}
\end{eqnarray} 
(notice however the difference with the expression proposed in 
\cite{liu})
which brings the Eq. (\ref{req1}) back into the form of the
regular NLS equation, without the external potential, in which
case the condition of Eq. (\ref{req8}) applies.

Hence, the growth terms will now be $\sim \exp(i (Q \xi-\Omega t))$
with $\Omega = \Omega_r + i \nu$ (when the instability condition is satisfied;
$\Omega_r=2 q Q$, cf. Eq. (\ref{req8})),
and hence the instability will be developing  according to the
spatiotemporal form:
\begin{eqnarray}
u \sim \exp \left(i Q (x+\alpha t^2)  - i \Omega_r t + \nu t -i \alpha x t 
-\frac{1}{3} i \alpha^2 t^3 \right).
\label{req10}
\end{eqnarray}
%Notice, however, that now the wavenumber corresponds to the
%new spatial variable $\xi$.

\subsection{Quadratic Potential}

The quadratic potential of Eq. (\ref{req3}) is clearly the most
physically relevant example of an external potential in the BEC
case, given the harmonic confinement of the atoms by the
experimentally used magnetic traps \cite{review}. 
In particular, to examine the MI related properties in this case,
we will use a lens-type transformation \cite{sulem} 
of the form:
\begin{equation}
u(x,t)=\ell^{-1} \exp(i f(t) x^2)v(\zeta,\tau)
\label{req11}
\end{equation}
where $f(t)$ is a real function of time, $\zeta=x/\ell(t)$ and $\tau=\tau(t)$. 
%Eq. (\ref{req2}) then becomes
%\begin{equation}
%i \tau_{t} \sigma \frac{\ell_{\tau}}{\ell} v-i v_{\zeta} x  \frac{\ell_{\tau} \tau_{t}}
%{\ell^2}+4 i x \frac{f}{\ell} v_{\zeta} + 2 i f v=0
%\label{req13}
%\end{equation}
To preserve the scaling we choose \cite{sulem,skk}
\begin{eqnarray}
\tau_{t}=1/\ell^2
\label{req13b}
\end{eqnarray}
To satisfy the resulting equations, we then demand that:
\begin{eqnarray}
-f_{t}=4f^2+k(t)
\label{req14a}
\\
-\ell_{\tau}/\ell+4 f \ell^2=0.
\label{req14b}
\end{eqnarray}
Taking into account (\ref{req13b}) the last equation can be solved:
\begin{equation}
\label{req14c}
\ell(t)=\ell(0)\exp\left(4\int_0^tf(s)ds\right)
\end{equation}
what reduces the problem of finding time dependence of the parameters to solution of Eq. (\ref{req14a}).

Upon the above conditions, the equation for $v(\zeta,\tau)$ becomes
\begin{equation}
iv_{\tau}+v_{\zeta\zeta}+|v|^2 v-2 i\lambda v =0;  
\label{req15}
\end{equation}
where 
\begin{equation}
f \ell^2=\lambda,
\label{req16}
\end{equation}
and generically $\lambda$ is real and depends on time. Thus we retrieve 
NLS with an additional term, which represents
either  growth (if $\lambda>0$) or dissipation (if $\lambda<0$). 

Eq. (\ref{req15}) allows an explicit solution when 
$v \equiv v(\tau)$ (i.e., a spatially homogeneous solution). The latter
is of the form
$$
v = \frac{A_0}{\ell(t)}  
    \exp \left(i f(t) x^2 + i q \frac{x}{\ell(t)} - i q^2 \tau(t) \right) 
\times
$$
\begin{eqnarray}
\exp \left( \Lambda(\tau)+ i \theta_0 + i A_0^2 \int_0^{\tau} \exp(2 \Lambda(s)) ds \right),
\label{exact}
\end{eqnarray}
where $\Lambda(\tau)= 4 \int_0^{\tau} \lambda(s) ds$, 
and $A_0$ and $\theta_0$ are arbitrary real constants.

A particularly interesting case is that of $\lambda$ constant. 
Then from the system of equations
(\ref{req13b})-(\ref{req14b})
and (\ref{req15}) 
it follows that $k$ must have a specific form. $f$,$\ell$ and 
$\tau$ can then be determined accordingly. 
In fact, the system (\ref{req13b})-(\ref{req14b})
and (\ref{req15})  with $\lambda$ constant has as its
solution
\begin{eqnarray}
k(t)&=&(t+t^{*})^{-2}/16
\label{req18a}
\\
f(t)&=&(t+t^{*})^{-1}/8
\label{req18b}
\\
\ell(t)&=&2 \sqrt{2 \lambda} \sqrt{t+t^{*}}
\label{req18c}
\\
\tau(t)&=&\ln (\frac{t+t^{*}}{t^{*}})/8 \lambda.
\label{req18d}
\end{eqnarray}
Notice once again, per the assumption of 
an imaginary phase in the exponential of Eq. (\ref{req11}), that
our considerations are valid only for $\lambda \in R$. $t^{\star}$ 
in the above equations is an arbitrary constant that essentially 
determines the ``width'' of the trap at time $t=0$ according 
to Eq. (\ref{req18a}).
% and
%\begin{eqnarray}
%\ell(t)&=&2 \sqrt{2 |\lambda|} \exp[i (\pi/4+\kappa \pi)]\sqrt{t+t^{*}}
%\label{req19a}
%\\
%\tau(t)&=&- i\ln(\frac{t+t^{*}}{t^{*}})/8 |\lambda|,
%\label{req19b}
%\end{eqnarray}
%where $\kappa$ is any integer, for  $\lambda$ purely imaginary.
%In the latter case, the transformation 
%$v \rightarrow w(\zeta,\tau) \exp(-i 4 (i \lambda) \tau)$
%results in the regular NLS for $w$ and hence the modulational condition
%remains unchanged. In this case, the unstable eigenmodes grow
%as $w \sim \exp(i (Q \zeta-\Omega_r \tau) + \nu \tau)$, 
%when $\Omega = \Omega_r + i \nu$. 

%If $\lambda$ is real, then similar results are obtained. In fact,
In this case
the 
modulational condition remains unchanged, but now $\omega$ satisfies the 
dispersion 
relation $\omega=q^{2}-\phi^{2}+2 \imath \lambda$, so the growth 
(if $\lambda>0$) or 
dissipation (if $\lambda<0$) is inherent in 
equation (\ref{req5}). Moreover, all the terms are 
modulated by the constant growth (or decay) rate $\exp(2 \lambda \tau)$, and 
the instability (when present) will be developing according to the form 
$v\sim \exp \left(i (Q \zeta-\Omega_r \tau)  + (\nu+2 \lambda) \tau\right)$ 
with 
$\Omega=\Omega_r + i \nu$. $\Omega_r=2 q Q$.

If $k=k(t)$ is not given by Eq. (\ref{req18a}), 
then $\lambda$ must be time dependent
(e.g., $\lambda \equiv \lambda(t)$).
Here one cannot directly perform the
MI analysis. However, still in this case, we have converted the explicit
spatial dependence into an explicit temporal dependence.
An important conclusion that stems from this transformation is that
the harmonic potential, viewed in the appropriate frame 
(of Eq. (\ref{req11})) can be considered as a form of growth (or dissipation,
depending on the sign of $\lambda$ for $\lambda \in R$) 
term.

\section{Conclusions}

In this brief communication, we have examined the problem of modulational
instabilities of plane waves in the context of Gross-Pittaevskii equations
with an external (linear or quadratic) potential. Both of these cases are
directly relevant to current experimental realizations of Bose Einstein
condensates. 

It was found that the linear (also integrable) case can be reverted to
the original NLS setting (wherein the MI conditions are well-known) by
an appropriate (Tappert) transformation. This transformation was used
to develop the form of the growth of unstable wavenumbers. Notice, however,
that these are the wavenumbers of the expansion in a novel spatial coordinate
which is essentially the spatial variable in a frame with constant 
acceleration. Hence, in this case, in the frame moving with a constant
acceleration (induced by the constant force), the problem resumes its
original NLS format and the MI conditions and resulting growth can be
obtained in that frame and then restored in the original frame.

For the case of the quadratic potential, a lens transformation was
used to cast the problem in a rescaled space and time frame (in a
way very reminiscent of the scaling in problems related to focusing
\cite{sulem,skk}). In this rescaled frame, the external potential
can be viewed as a form of external growth. For specific forms
of temporal dependence of the prefactor of the harmonic term (e.g., 
for appropriate, non-autonomous quadratic potentials), the resulting
prefactor is constant. In such a context once again the MI analysis
can be carried through completely, producing similar
conditions, but now in the new dynamically rescaled frame/variables
(which can be appropriately re-cast in the original variables).

It would be interesting to examine if similar considerations 
can be generalized to the case of multiple dimensions. Furthermore,
in the current scheme of things it seems that the cases of different
forms of the potentials need to be treated in different ways (which
can be understood in terms of the different physical effects that
they represent). Nonethless, it would be very useful, if a general
formulation could be developed that could be applied independently
of the form of $V(x)$, having as special case limits, the potential
forms presented herein.

Finally, it would be worth exploring whether the explicitly demonstrated
as modulationally unstable settings presented herein can be used
as a means (i.e., initial condition) for directly producing 
solitary (matter) wave structures
in BECs in an alternative fashion to the ones currently used in 
BEC experiments.

Prof. D.J.
Frantzeskakis is gratefully acknowledged for a critical reading 
of the manuscript and his suggestions.
PGK gratefully acknowledges support from a University of Massachusetts
Faculty Research Grant, from the Clay Foundation through a Special 
Project Prize Fellowship and from the NSF through DMS-0204585.
VVK gratefully acknowledges support 
from the European grant COSYC n.o {HPRN-CT}-2000-00158.


\begin{thebibliography}{0}

\bibitem{review}  F. Dalfovo, S. Giorgini, L. P. Pitaevskii, and S.
Stringari, %Theory of Bose-Einstein condensation in trapped gases, 
Rev. Mod. Phys. {\bf 71}, 463%-521 
(1999). 
%\bibitem{Kett}  M.H. Anderson {\it et al}, Science {\bf 269}, 198 (1995);
%K.B. Davis {\it et al}, Phys. Rev. Lett. {\bf 75}, 3969 (1995); C.C. Bradley 
%{\it et al}, Phys. Rev. Lett. {\bf 75}, 1687 (1995).

\bibitem{bright}  K.E. Strecker {\em et al.}, Nature {\bf 417}, 150 (2002);
L. Khaykovich {\em et al.}, Science {\bf 296}, 1290 (2002).

\bibitem{dark}  S. Burger {\it et al.}, Phys. Rev. Lett. {\bf 83}, 5198
(1999). 

\bibitem{dexp2}  J. Denschlag {\it et al.}, Science {\bf 287}, 97 (2000).

\bibitem{nsbec}  B.P. Anderson {\it et al.}, Phys. Rev. Lett. {\bf 86}, 2926
(2001).

\bibitem{vortex}  M.R. Matthews {\it et al.}, Phys. Rev. Lett. {\bf 83},
2498 (1999); K.W. Madison {\it et al.} Phys. Rev. Lett. {\bf 84}, 806
(2000); S. Inouye {\em et al.}, Phys. Rev. Lett.{\bf {87}}, 080402 (2001).

\bibitem{vl}  J.R. Abo-Shaeer {\it et al}, Science {\bf 292}, 476 (2001);
J.R. Abo-Shaeer, C. Raman and W. Ketterle, Phys. Rev. Lett. {\bf {88}},
070409 (2002); P. Engels {\it et al}, Phys. Rev. Lett. {\bf 89}, 100403
(2002).

\bibitem{sulem} C. Sulem and P.L. Sulem,
\newblock {\it The Nonlinear Schr{\"o}dinger Equation},
Springer-Verlag (New York, 1999).


\bibitem{benjamin67} T.B. Benjamin and J.E. Feir, 
J. Fluid. Mech. {\bf 27}, 417 (1967).

\bibitem{ostrovskii69} L.A. Ostrovskii, 
Sov. Phys. JETP {\bf 24}, 797 (1969).

\bibitem{taniuti68} T. Taniuti and H. Washimi, 
Phys. Rev. Lett. {\bf 21}, 209 (1968); A. Hasegawa, 
Phys. Rev. Lett. {\bf 24}, 1165 (1970).

\bibitem{konotop} V.V. Konotop and M. Salerno,
\newblock Phys. Rev. A {\bf 65}, 021602(R) (2002).

\bibitem{smerzi}  A. Smerzi, A. Trombettoni, P.G. Kevrekidis,
and  A.R. Bishop, cond-mat/0207172 (Phys. Rev. Lett., in press).

\bibitem{cattal} F.S. Cataliotti {\it et al.}, 
cond-mat/0207139.

\bibitem{kasevich} M. Kasevich and A. Tuchman 
(private communication).


\bibitem{kasevich1} B. P. Anderson and M. A. Kasevich, Science, 
{\bf 282}, 1686 (1998).

 

\bibitem{liu} H.-H. Chen and C.-S. Liu, Phys. Rev. Lett.,
{\bf 37}, 693 (1976).

\bibitem{skk} see e.g., S.I. Siettos, I.G. Kevrekidis and
P.G. Kevrekidis, nlin.PS/0204030 and references therein.


\end{thebibliography}
\end{document}